\def\year{2020}\relax

\documentclass[letterpaper]{article}
\usepackage{aaai20}  % DO NOT CHANGE THIS
\usepackage{times}  % DO NOT CHANGE THIS
\usepackage{helvet} % DO NOT CHANGE THIS
\usepackage{courier}  % DO NOT CHANGE THIS
\usepackage[hyphens]{url}  % DO NOT CHANGE THIS
\usepackage{graphicx} % DO NOT CHANGE THIS
\usepackage[utf8]{inputenc}
\usepackage{mathtools}
\usepackage[ruled,linesnumbered]{algorithm2e}
\usepackage{epsfig}
\usepackage{epstopdf}
\usepackage{amsmath}
\usepackage{subfig}
\usepackage[noend]{algorithmic}
\usepackage{amssymb}
\usepackage{textcomp}
\usepackage{mathtools}
\usepackage{rotating}
\usepackage{placeins}

\DeclarePairedDelimiter\ceil{\lceil}{\rceil}
\DeclarePairedDelimiter\floor{\lfloor}{\rfloor}

\newlength\myindent
\newcommand\bindent{%
	\begingroup
	\setlength{\itemindent}{\myindent}
	\addtolength{\algorithmicindent}{\myindent}
}
\newcommand\eindent{\endgroup}

\newcommand{\squishlist}[1][$\bullet$]
{
	\begin{list}{#1}
		{ 
			\setlength{\leftmargin}{1em}
			\setlength{\labelwidth}{1em}
			\setlength{\labelsep}{0.5em}
		} 
	}
	
	\newcommand{\squishend}{\end{list}}

\newcommand{\squishenum}
{
	\begin{itemize}
		{ 
			\setlength{\leftmargin}{1em}
			\setlength{\labelwidth}{1em}
			\setlength{\labelsep}{0.5em}
		} 
	}
	
	\newcommand{\squishenumend}{\end{itemize}}
\title{Neural Approximate Dynamic Programming for On-Demand Ride-Pooling}
\author{  \Large \textbf{Sanket Shah, Meghna Lowalekar, Pradeep Varakantham}\\
	\small School of Information Systems, Singapore Management University \\
	\footnotesize sankets@smu.edu.sg, meghnal.2015@phdcs.smu.edu.sg, pradeepv@smu.edu.sg
}

\newcommand\Tstrut{\rule{0pt}{2.6ex}}         % = `top' strut
\newcommand\Bstrut{\rule[-0.9ex]{0pt}{0pt}}   % = `bottom' strut

\nocopyright

\begin{document}

\maketitle

\begin{abstract}
On-demand ride-pooling (e.g., UberPool, LyftLine, GrabShare) has recently become popular because of its ability to lower costs for passengers while simultaneously increasing revenue for drivers and aggregation companies (e.g., Uber). Unlike in Taxi on Demand (ToD) services -- where a vehicle is assigned one passenger at a time -- in on-demand ride-pooling, each (possibly partially filled) vehicle must be assigned a group of passenger requests with multiple different origin and destination pairs in such a way that quality constraints are not violated. To ensure near real-time response, existing solutions to the real-time ride-pooling problem are myopic in that they optimise the objective (e.g., maximise the number of passengers served) for the current time step without considering the effect such an assignment could have on feasible assignments in future time steps. However, considering the future effects of an assignment that already has to consider what combinations of passenger requests can be assigned to vehicles adds an extra layer of combinatorial complexity on top of the already challenging problem of considering future effects in the ToD case. 

A popular approach that addresses the limitations of myopic assignments in ToD problems is Approximate Dynamic Programming (ADP). Existing ADP methods for ToD can only handle Linear Program (LP) based assignments as the value update relies on dual values from the LP. The assignment problem in ride pooling requires an Integer Linear Program (ILP) that has bad LP relaxations. Therefore, our key technical contribution is in providing a general ADP method that can learn from the ILP based assignment found in ride-pooling. Additionally, we handle the extra combinatorial complexity from combinations of passenger requests by using a Neural Network based approximate value function and show a connection to Deep Reinforcement Learning that allows us to learn this value-function with increased stability and sample-efficiency. We show that our approach easily outperforms leading approaches for on-demand ride-pooling on a real-world dataset by up to 16\%, a significant improvement in city-scale transportation problems.
\end{abstract}

\section{Introduction}

On-demand ride-pooling, exemplified by UberPool, LyftLine, GrabShare, etc., has become hugely popular in major cities with 20\% of all Uber trips coming from their ride-pool offering UberPool \cite{uberpool-stats,rideshare-stats}. Apart from reducing emissions and traffic congestion compared to Taxi/car on-Demand (ToD) services (e.g., UberX, Lyft), it benefits all the stakeholders involved: (a) Individual passengers have reduced costs as these are shared by overlapping passengers; (b) Vehicles make more money per trip as multiple passengers (or passenger groups) are present; (c) For the centralized entity (like Uber, Lyft etc.) more customer requests can be satisfied with the same number of vehicles.

Underlying these on-demand ride-pooling services is the Ride-Pool Matching Problem (RMP) \cite{alonso2017demand,aaairide,lowalekarVJ19}. The objective in the RMP is to assign groups of user requests to vehicles that can serve them, online, subject to predefined quality constraints (e.g., the detour delay for a customer cannot be more than 10 minutes) in such a way that a quality metric is maximised (e.g., revenue). The RMP reduces to the taxi/car-on-demand (ToD) problem when the capacity, i.e., the maximum number of simultaneous passengers (with different origin and destination pairs) that can be served by a vehicle, is 1. In this paper, we consider the most general version of the RMP, in which batches of requests are assigned to vehicles of arbitrary capacity, and present a solution that can scale to practical use-cases involving thousands of locations and vehicles.

Past research in solving RMP problems can be categorized along four threads. Past work along the first thread employs traditional planning approaches to model the RMP as an optimisation problem \cite{Ropke2009,ritzinger2016survey,parragh2008survey}. The problem with this class of approaches is that they don't scale to on-demand city-scale scenarios. The second thread consists of approaches that make the best greedy assignments \cite{ma2013t,tong2018unified,huang2014large,lowalekarVJ19,alonso2017demand}. While these scale well, they are myopic and, as a result, do not consider the impact of a given assignment on future assignments. The third thread, consists of approaches that use Reinforcement Learning (RL) to address the myopia associated with approaches from the second category for the ToD problem \cite{xu2018large,Lin:2018,li2019efficient,wang2018deep,verma2017augmenting}. Past RL work for the ToD problem cannot be extended to solve the RMP, however, because it relies heavily on the assumption that vehicles can only serve one passenger at a time.

Lastly, there has been work in operations research that uses the Approximate Dynamic Programming (ADP) framework to solve the ToD problem~\cite{powell2007approximate} and a special case of a capacity-2 RMP~\cite{Yu2019}. In ADP, matching is performed using a learned value function that estimates the future value of performing a certain matching. There are two major reasons for why past ADP approaches for solving the ToD problem cannot immediately be applied to the RMP, however. Firstly, the value function approximation in past work is linear and its update relies heavily on the assumption that matching can be modelled as a Linear Program (LP); this does not hold for the RMP with arbitrary vehicle capacities. Secondly, in the RMP, passengers may be assigned to a partially filled vehicle at each time step. This results in a complex state space for each vehicle that is combinatorial (combinations of requests already assigned) in the vehicle capacity (more details in Section~\ref{sect:challenges}).

We make three key contributions in this paper. First, we formulate the arbitrary capacity RMP problem as an Approximate Dynamic Programming problem. Second, we propose Neural ADP (NeurADP), a general ADP method that can learn value functions (approximated using Neural Networks) from ILP based assignment problems. Finally, we bring together techniques from Deep Q-Networks (DQN)~\cite{mnih2015human} to improve the stability and scalability of NeurADP.

In the experiments, we compare our approach to two leading approaches for the RMP on a real-world dataset \cite{yellowtaxi}. Compared to a baseline approach proposed by \cite{alonso2017demand}, we show that our approach serves up to 16\% more seen requests across different parameter settings. This translates to a relative improvement of 40\% over the baseline.

\section{Background: Approximate Dynamic Programming (ADP)}
\label{sect:bg}

ADP is a framework based on the Markov Decision Problem (MDP) model for tackling large multi-period stochastic fleet optimisation problems~\cite{powell2007approximate} such as ToD. The problem is formulated using the tuple $\left<S,A,\xi,T,O\right>$:
\squishlist
\item[$S$:] denotes the system state with $s_{t}$ denoting the state of system at decision epoch $t$.
\item[$A$:] denotes the set of all possible actions~\footnote{We use action and decision interchangeably in the paper.} (which satisfy the constraints on the action space) with $A_{t}$ denoting the set of possible actions at decision epoch $t$. $a_{t} \in A_{t}$ is used to denote an action at decision epoch $t$.
\item[$\xi$:] denotes the exogenous information -- the source of randomness in the system. For instance, this would correspond to demand in ToD problems. $\xi_t$ denotes the exogenous information (e.g., demand) at time $t$. 
\item[$T$:] denotes the transition function which describes how the system state evolves over time.
\item[$O$:] denotes the objective function with $o_{t}(s_{t},a_{t})$ denoting the value obtained on applying action $a_{t}$ on state $s_{t}$.
\squishend

In an MDP, system evolution happens as $(s_0, a_0, s_1, a_1, s_2, ....)$. However, in an ADP, the evolution happens as $(s_{0},a_{0},s_{0}^{a},\xi_{1},s_{1},a_{1},s_{1}^{a},\cdots,s_{t},a_{t},s_{t}^{a},\cdots)$, where $s_{t}$ denotes the pre-decision state at decision epoch $t$ and $s_{t}^{a}$~\footnote{Here $a$ is just used to indicate that it is post decision state and it does not correspond to any specific action.} denotes the post-decision state~\cite{powell2007approximate}. The transition from state $s_{t}$ to $s_{t+1}$ depends on the action vector $a_{t}$ and the exogenous information $\xi_{t+1}$.  Therefore,
$$s_{t+1} = T(s_{t},a_{t},\xi_{t+1}) $$ 
Using post-decision state, this transition can be written as
$$s_{t}^{a} = T^{a}(s_{t},a_{t}); s_{t+1} = T^{\xi}(s_{t}^{a},\xi_{t+1})$$

Let $V(s_{t})$ denotes the value of being in state $s_{t}$ at decision epoch $t$, then using Bellman equation we get
$$ V(s_{t}) = \max_{a_{t} \in A_{t}} (O(s_{t},a_{t}) + \gamma \mathbb{E}[V(s_{t+1})|s_{t},a_{t},\xi_{t+1}])$$
where $\gamma$ is the discount factor. Using post-decision state, this expression can be broken down into two parts:
{\small 
\begin{align}
V(s_{t}) &= \max_{a_{t} \in A_{t}} (O(s_{t},a_{t}) + \gamma V^{a}(s_{t}^{a})) \label{eqn:1}\\
V^{a}(s_{t}^{a}) &= \mathbb{E}[V(s_{t+1})|s_{t}^{a},\xi_{t+1}] \label{eqn:2}
\end{align}
}
The advantage of this decomposition is that Equation~\ref{eqn:1} can be solved using an LP in fleet optimisation problems. 
The basic idea in any ADP algorithm is to define a value function approximation around post-decision state, $V^{a}(s_{t}^{a})$) and to update it by stepping forward through time using sample realizations of exogenous information (i.e. demand in fleet optimisation that is typically observed in data). Please refer to Powell~\cite{powell2007approximate} for more details. 

\begin{figure*}[t]
	\centering
	\includegraphics[width=\textwidth,height=3.75in]{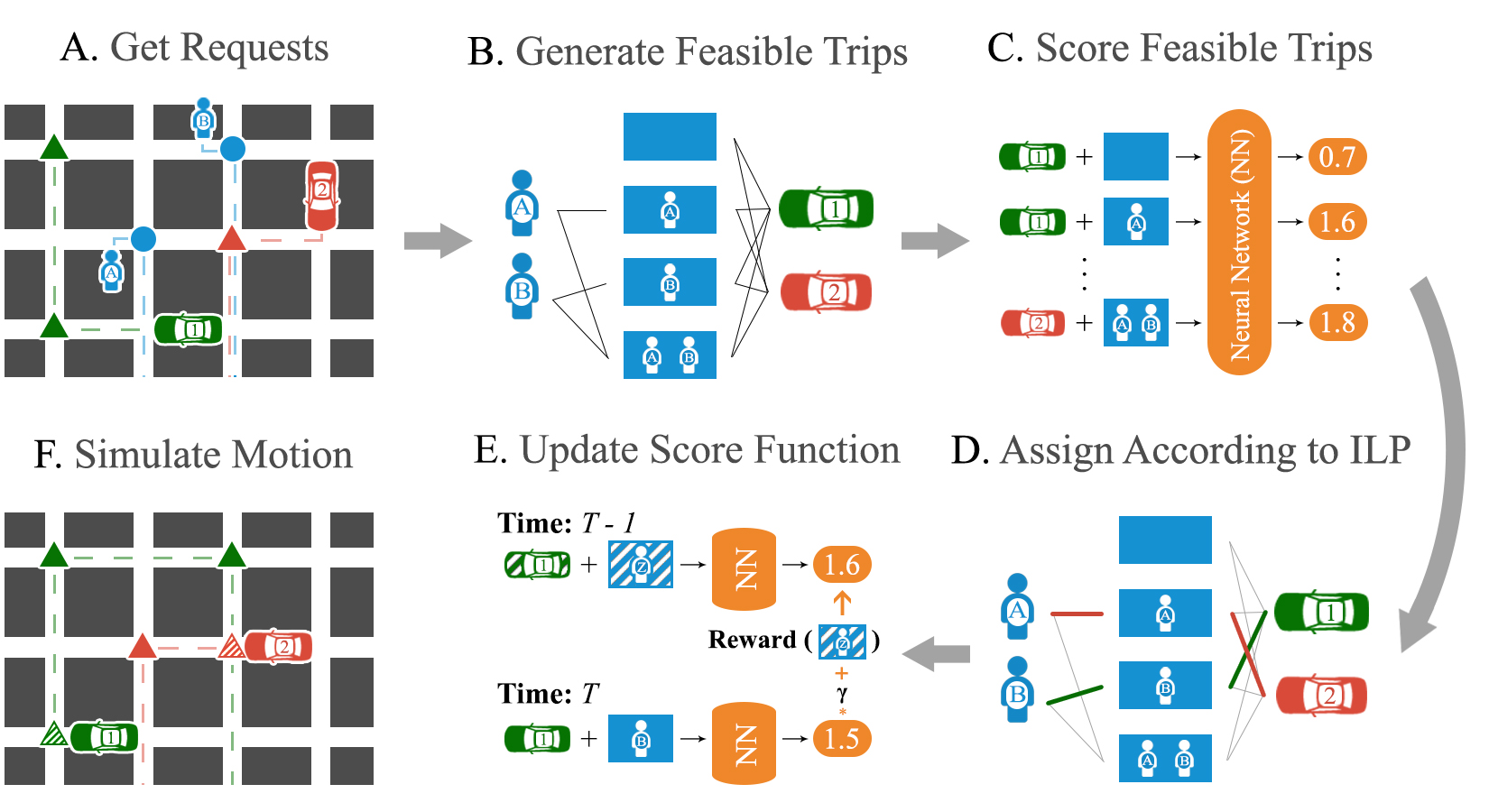}
	
	\caption{\textbf{Schematic outlining our overall approach.} We start with a hypothetical $\cal{G}$, $\cal{U}$ and $\cal{R}$ in \textbf{(A)}. The grid represents a road network. The blue people and circles correspond to user requests and the nearest street intersection that they're mapped to respectively. The blue dotted lines represent the shortest path between the pick-up and drop-off points of a request. The red and green triangles correspond to existing pick-up/drop-off points for the red and green vehicles respectively. The dotted lines describe their current trajectory. In \textbf{(B)} we map the requests and their combinations to vehicles that can serve them under the constraints defined by $\cal{D}$ to create feasible actions using the approach presented in \cite{alonso2017demand}. In \textbf{(C)}, we score each of these feasible actions using our Neural Network Value Function. In \textbf{(D)}, we create a mapping of requests to vehicles that maximises the sum of scores generated in (C) using the Integer Linear Program (ILP) in Table \ref{table:opt}. In \textbf{(E)}, we use this final mapping to update the score function (Section \ref{sect:tdlearning}). In \textbf{(F)}, we simulate the motion of vehicles until the next epoch either based on their current trajectories or a re-balancing strategy. This process then repeats for the next decision epoch.}
	
	\label{fig:approach}
\end{figure*}

\section{Ride-pool Matching Problem (RMP)}
In this problem, we consider a fleet of vehicles/resources ${\cal R}$ with random initial locations, travelling on a predefined road network ${\cal G}$. Passengers that want to travel from one location to another send requests to a central entity that collects these requests over a time-window called the decision epoch $\Delta$. The goal of the RMP is to match these collected requests $U^t$ to empty or partially filled vehicles that can serve them such that an objective ${\cal O}$ is maximised subject to constraints on the delay ${\cal D}$. These delay constraints ${\cal D}$ are important because customers are only willing to trade being delayed in exchange for a reduced fare up to a point. In this paper, we consider the objective ${\cal O}$ to be the number of requests served.

We provide a formal definition for the RMP using the tuple $\left<{\cal G}, {\cal U}, {\cal R}, {\cal D}, \Delta, {\cal O}\right>$:
\squishlist
    \item[${\cal G}$:] $({\cal L},{\cal E})$ is a weighted graph representing the road network. Along the lines of \cite{alonso2017demand}, ${\cal L}$ denotes the set of street intersections and ${\cal E}$ defines the adjacency of these intersections. Here, the weights correspond to the travel time for a road segment. We assume that vehicles only pick up and drop people off at intersections.
    \item[${\cal U}$:] denotes the set of user requests. ${{\cal U} = \times_{t}\: {{\cal U}_{t}}}$ is the combination of requests that we observe at each decision epoch $t$. Each request $u_t^{j} \in {\cal U}_{t}$ is represented by the tuple: $\left<o^j_t, e^j_t, t\right>$, where $o^j_t, e^j_t \in {\cal L}$ denote the origin and destination and $t$ denotes the arrival epoch of the request. 
    \item[${\cal R}$:] denotes the set of resources/vehicles. Each element $i \in {\cal R}$ is represented by the tuple $\left<c^i, p^i, L^i\right>$. $c^i$ denotes the capacity of the vehicle, i.e., the maximum number of passengers it can carry simultaneously, $p^i$ its current position and $L^i$ the ordered list of locations that the vehicle should visit next to satisfy the requests currently assigned to it.
    \item[${\cal D}$:] $\{\tau, \lambda\}$ denotes the set of constraints on delay. $\tau$ denotes the maximum allowed pick-up delay which is the difference between the arrival time of a request and the time at which a vehicle picks the user up. $\lambda$ denotes the maximum allowed detour delay which is the difference between the time at which the user arrived at their destination in a shared cab and the time at which they would have arrived if they had taken a single-passenger cab.     
	\item[$\Delta$:] denotes the decision epoch duration.
    \item[${\cal O}$:] represents the objective, with ${\cal O}^{i}_{t}$ denoting the value obtained by serving request $i$ at decision epoch $t$. The goal of the online assignment problem is to maximize the overall objective over a given time horizon, $T$.
\squishend 

\section{NeurADP: Neural Approximate Dynamic Programming} 
\label{sect:adp}
Figure \ref{fig:approach} represents the overall framework used for solving the RMP. As shown in the figure, the framework executes 6 steps at each decision epoch to assign incoming user requests to available vehicles. Existing myopic approaches only execute steps (A), (B), (D) and (F). The crucial steps (C) and (E) help in maximising the expected long-term value of serving a request rather than its immediate value. To learn this long-term value, we model the RMP problem using ADP and use deep neural networks to learn the value functions of post-decision states. 

In this section, we first indicate key challenges that preclude direct application of existing ADP methods. We next provide the ADP model for the RMP problem and describe our contributions in using neural function approximations for scalable and effective policies in RMP. 

\subsection{Departure From Past Work} \label{sect:challenges}
Approximate Dynamic Programming has been used to model many different transportation problems such as fleet management~\cite{simao2009approximate}, ambulance allocation~\cite{maxwell2010approximate} etc. While we also model our RMP problem using ADP, we cannot use the solutions from past work for the following reasons:
\squishlist
\item[1.] \textbf{\textit{Non-trivial generation of feasible actions:}}
In using ADP to solve the ToD problem, the action for a single empty vehicle is to match a single request. Computing the feasible set of requests for a vehicle is a straightforward and the best action for all vehicles together can then be computed by solving a Linear Program (LP). In the case of the RMP, multiple requests can be assigned to a single empty or partially filled vehicle. Generating the set of feasible actions, in this case, is complex and real-time solutions to this problem have been the key challenge in literature on myopic solutions to ride-pooling. In this paper, we use the approach proposed by~\cite{alonso2017demand} to generate feasible actions for a single vehicle (Section \ref{sect:adp}) and then use an Integer Linear Program (ILP) to choose the best action (Table \ref{table:opt}) over all vehicles.

\item[2.] \textbf{\textit{Inability to use LP-duals to update the value function:}}
Past work in ADP for ToD~\cite{simao2009approximate} uses the dual values of the matching LP to update the parameters of their value function approximation. However, choosing the best action in the RMP requires solving an Integer Linear Program (ILP) that has bad LP-relaxations. As a result, we cannot use duals to update our value function. Instead, we show the connection between ADP and Reinforcement Learning (RL), and use the more general Bellman update used in RL to update the value function (Section \ref{sect:tdlearning}).

\item[3.] \textbf{\textit{Curse of Dimensionality:}} Past work in ADP for transportation problems addresses the curse of dimensionality by considering the value function to be dependent on a small set of hand-crafted attributes (e.g., aggregated number of vehicles in each location) rather than on the states of a large number of vehicles. Hand-crafting of state attributes is domain-specific and is incredibly challenging for a complex problem like RMP, where aggregation of vehicles is not a feasible attribute (as each vehicle can have different number of passengers going to multiple different locations). Instead, we use a Neural Network based value function to automatically learn a compact low dimensional representation of the large state space. 

\item[4.] \textbf{\textit{Incorporating Neural Network value functions into the optimisation problem:}}
Past work in ADP for ToD uses linear or piece-wise linear value function approximations that allow for the value function to be easily integrated into the matching LP. Non-linear value functions (such as neural networks) cannot be integrated in this way, however, as they would make the overall optimisation program non-linear. In Section \ref{sect:approx}), we address this issue by using a two-step decomposition of the value function that allows it to be efficiently integrated into the ILP as constants.

\item[5.] \textbf{\textit{Challenges of learning a Neural Network value function:}}
In Deep Reinforcement Learning literature~\cite{mnih2015human}, it has been shown that naive approaches to approximating Neural Network value functions are unstable. Additionally, training them requires millions of samples. To address this, we propose a combination of methodological and practical solutions in Section \ref{sect:learningchallenges}.
\squishend
The combination of using a Neural Network value function (instead of linear approximations) and updating it with a more general Bellman update (instead of LP-duals) represents a general alternative to past ADP approaches that we term Neural ADP (NeurADP). 

\subsection{Approximate Dynamic Programming Model for the RMP}
We model the RMP by instantiating the tuple in Section~\ref{sect:bg}. 

\squishlist
\item [$S$:] The state of the system is represented as $s_{t}=(r_{t},u_{t})$ where $r_{t}$ is the state of all vehicles and $u_{t}$ is contains all the requests waiting to be served. A vehicle $r \in {\cal R}$ at decision epoch $t$ is described by a vector $r_{t}^{i}=(p^{i},t,L^{i})$ which represents its current trajectory. Specifically, it captures the current location ($p^i$), time($t$) and an ordered list of future locations (along with the cut-off time by which each must be visited) that the vehicle has to visit ($L^i$) to satisfy the currently assigned requests. Each user request $j$ at decision epoch $t$ is represented using vector $u_{t}^{j}= (o^{j},e^{j})$ which captures its origin and destination. \\

\item[$A$:] For each vehicle, the action is to assign a group of users from the set ${\cal U}_{t}$ to it. These actions should satisfy:
\squishlist
\item[1.] \textbf{\textit{Constraints at the vehicle level}} - satisfying delay constraints ${\cal D}$ and vehicle capacity constraints
\item[2.] \textbf{\textit{Constraints at the system level}} - Each request is assigned to at most one vehicle. 
\squishend
~\\ 
\textbf{Handling exponential action space:} To reduce the complexity, feasible actions are generated in two steps. In the first step, we handle vehicle-level constraints by generating a set of feasible actions (groups of users) for each vehicle. To do this efficiently, we first generate an RTV (Request, Trip, Vehicle) graph using the algorithm by Alonso et.al.~\cite{alonso2017demand}. Along with feasible actions, this generation process also provides the routes that the vehicle should take to complete each action. We use ${\cal F}_{t}^{i}$ to denote the set of feasible actions generated for vehicle $i$ at decision epoch $t$. 
{\small
 \begin{align}
            {\cal F}_{t}^{i} &= \{f^i \,|\,f^i\in \cup_{c'=1}^{c^i}\, [{\cal U}]^{c'}, PickUpDelay(f^i,i) \leq \tau, \nonumber \\
		& \hspace{1in} \,DetourDelay(f^i,i) \leq \lambda \} \nonumber
        \end{align}}
To ensure that system-level constraints are satisfied, we solve an ILP that considers all vehicles and requests together. Let $a_{t}^{i,f}$ denote that vehicle $i$ takes action $f$ at decision epoch $t$. Then, the decision variables $a_{t}^{i,f}$ need to satisfy following constraints:
{\small
\begin{align}
\quad & \sum_{f \in {\cal F}_{t}^{i}} a_{t}^{i,f} = 1 ::: \forall i \in {\cal R} \label{cons:a1}\\
& \sum_{i \in {\cal R}} \sum_{f \in {\cal F}_{t}^{i};j \in f} a_{t}^{i,f} \leq  1 ::: \forall j \in {\cal U}_{t} \label{cons:a2}\\
& a_{t}^{i,f} \in \{0,1\} \label{cons:a3} ::: \forall i,f
\end{align}
}

Constraint \eqref{cons:a1} ensures that each vehicle is assigned a single action and constraint \eqref{cons:a2} ensures that each request is a part of, at most, one action. Together, they ensure that a request can be mapped to at most one vehicle.

We use $A_{t}$ denote the set of all actions that satisfy both individual and system-level constraints at time $t$ and $a_{t} \in A_{t}$ to denote a feasible action in this set\footnote{At every time step, we augment $A_{t}$ with a \emph{null} action $a_\phi$. This allows a vehicle to continue along its trajectory without being assigned a passenger.}. \\

\item[$\xi$:]  As in previous work, exogenous information $\xi_t$ represents the user demand that arrives between time $t-1$ and $t$.  

\begin{table}
	\center
	\begin{tabular}{|r|}
		\hline
		
		\begin{minipage}{0.45\textwidth}
			\vspace{0.05in}
			\textbf{AssignmentILP(t):}
			{\small
				\begingroup
				\begin{align}
				\max \quad & \sum_{i} \sum_{f \in {\cal F}_{t}^{i}} o_{t}^{i,f}*a_{t}^{i,f} + V^{i}(T^{i,a}(r_{t}^{i,a},f))*a_{t}^{i,f} \label{opt:objective}\\
				subject ~ to ~& \text{ Constraints \eqref{cons:a1} - \eqref{cons:a3}} \nonumber
				\end{align}
				
				\endgroup }
		\end{minipage} \\
		\hline
	\end{tabular}
	\caption{ Optimization Formulation for assignment of vehicles to feasible actions}
	\label{table:opt}
\end{table}

\item[$T$:] The transition function $T^{a}$ defines how the vehicle state changes after taking an action. In the case of the RMP, all user requests that are not assigned are lost~\cite{alonso2017demand}. Therefore, the user demand component of post-decision state will be empty, i.e., $u_{t}^{a} =\phi$.
{\small  
\begin{align}
T^{a}(s_{t},a_{t}) = r_{t}^{a} \label{eq:postdecision}
\end{align}}
Here, $r_t^a$ denotes the post decision state of the vehicles. We use $T^{i,a}(s_{t}^{i},a_{t}^{i}) = r_{t}^{i,a}$ to denote the transition of individual vehicles. At each decision epoch, based on the actions taken, $(p^{i},t,L^{i})\,\forall i$ are updated and are captured in $r_{t}^{i,a}$. Each vehicle has a fixed path corresponding to each action and as a result the transition above is deterministic. 

\item [$O$:] When vehicle $i$ takes a feasible action $f$ at decision epoch $t$, its contribution to the objective is $o_{t}^{i,f}$. For the objective of maximizing the number of requests served, $o_{t}^{i,f}$ is the number of requests that are part of a feasible action $f$ (0 for the null action $a_\phi$). The objective function at time $t$ is as follows: $$ o_{t}(s_{t},a_t) = \sum_{i \in {\cal R}}\sum_{f \in {\cal F}_{t}^{i}} o_{t}^{i,f}*a_{t}^{i,f} $$

\squishend

\begin{algorithm}[h]
	\caption{NeurADP ($N,T$)}
	{\small
		\begin{algorithmic}[1]
			\STATE{Initialize: replay memory $M$, Neural value function $V$\\(with random weights $\theta$)}
			\FOR{each episode $1 \leq n < N$}
			\STATE{Initialize the state $s_{0}^{n}$ by randomly positioning vehicles.}
			\STATE{Choose a sample path $\xi^{n}$}
			\FOR{each step $0 \leq t \leq T$}
			\STATE{Compute the feasible action set ${\cal F}_{t}$ based on $s_{t}^{n}$.}
			\STATE{Solve the ILP in Table \ref{table:opt} to get best action $a_{t}^{n}$. Add the Gaussian noise for exploration.}
			\STATE{Store ($r_{t}^{n},{\cal F}_{t}$) as an experience in $M$.}
			\IF{t \% updateFrequency == 0}
			\STATE{Sample a random mini-batch of experiences \\from $M$} 
			\FOR{each experience $e$}
			\STATE{Solve the ILP in Table \ref{table:opt} with the information \\from experience $e$ to get the objective value $y^{e}$}
			\FOR{each vehicle $i$}
			\STATE{Perform a gradient descent step on $(y^{e,i} - V(r_{t}^{i,n}))^2$ with respect to \\the network parameters $\theta$}
			\ENDFOR
			\ENDFOR
			\ENDIF
			\STATE{Update: {\small$s_{t}^{a,n}=T^{a}(s_{t}^{n},a_{t}^{n}),~s_{t+1}^{n}=T^{\xi}(s_{t}^{a,n},\xi_{t+1}^{n})$}}
			\ENDFOR
			\ENDFOR
		\end{algorithmic}
	}
	\label{alg:neuradp}
\end{algorithm}

\subsection{Value Function Decomposition}\label{sect:approx}
Non-linear value functions, unlike their linear counterparts, cannot be directly integrated into the matching ILP. One way to incorporate them is to evaluate the value function for all possible post-decision states and then add these values as constants. However, the number of post-decision states is exponential in the number of resources/vehicles.

To address this, we propose a two-step decomposition of our overall value function that converts it into a linear combination over individual value functions associated with each vehicle\footnote{As mentioned in equation \eqref{eq:postdecision}, the post-decision state only depends on the vehicle state. Therefore, $V(s_{t}^{a}) = V(r_{t}^{a})$.}:

\squishenum
\item \textbf{\textit{Decomposing joint value function based on individual vehicles' value functions:}}\label{sect:learning}
We use the fact that we have rewards associated with each vehicle to decompose the joint value function for all vehicles' rewards into a sum over the value function for each vehicle's rewards. The proof for this is straightforward and follows along the lines of \cite{russell2003q}.
$$V(r_{t}^{a}) = \sum_{i} V^{i}(r_{t}^{a})$$ 

\item \textbf{\textit{Approximation of individual vehicles' value functions:}}
We make the assumption that the long-term reward of a given vehicle is not significantly affected by the specific actions another vehicle makes in the current decision epoch. This makes sense because the long-term reward of a given vehicle is affected by the interaction between its trajectory and that of the other vehicles and, at a macro level, these do not change significantly in a single epoch. This assumption allows us to use the pre-decision, rather than post-decision, state of other vehicles. 
$$V^{i}(r_{t}^{a}) = V^{i}(\left<r_{t}^{i,a}, r_{t}^{\text{-}i,a}\right>) \approx V^{i}(\left<r_{t}^{i,a}, r_{t}^{\text{-}i}\right>)$$
Here, $\text{-}i$ refers to all vehicles that are not vehicle $i$. This step is crucial because the second term in the equation above $r_{t}^{\text{-}i}$ can now be seen as a constant that does not depend on the exponential post-decision state of all vehicles. 
\squishenumend

\noindent Therefore, the overall value function can be rewritten as:
$$V(r_{t}^{a}) = \sum_{i} V^{i}(\left<r_{t}^{i,a}, r_{t}^{\text{-}i}\right>)$$ 

We evaluate these individual $V^{i}$ values for all possible $r_{t}^{i,a}$ and then integrate the overall value function into the ILP in Table \ref{table:opt} as a linear function over these individual values. This reduces the number of evaluations of the non-linear value function from exponential to linear in the number of vehicles.

\subsection{Value Function Estimation for NeurADP}\label{sect:tdlearning}
To estimate the value function $V$ over the post-decision state, we use the Bellman equation (decomposed in the ADP as equation \eqref{eqn:1} and \eqref{eqn:2}) to iteratively update the parameters of the function approximation. In past work~\cite{simao2009approximate}, the parameters of a linear (or piece-wise linear) value function were updated in the direction of the gradient provided by the dual values at every step. Hence, the LP-duals removed the need to explicitly calculate the gradients in the case of a linear function approximation.

Given that we use a neural network function approximation and require an ILP (rather than an LP), we cannot use this approach. Instead, we use standard symbolic differentiation libraries~\cite{tensorflow2015-whitepaper} to explicitly calculate the gradients associated with individual parameters. We then update these parameters by trying to minimise the L2 distance between a one-step estimate of the return (from the Bellman equation) and the current estimate of the value function~\cite{mnih2015human}, as shown in Algorithm \ref{alg:neuradp}.

\subsection{Overcoming challenges in Neural Network Value Function Estimation}\label{sect:learningchallenges}
In this section, we describe how we mitigate the stability and scalability challenges associated with learning neural network value functions through a combination of methodological and practical methods. 
 
\subsubsection{Improving stability of Bellman updates:}
It has been shown in Deep Reinforcement Learning (DRL) literature that using standard on-policy methods to update Neural Network (NN) based value function approximations can lead to instability~\cite{mnih2015human}. This is because the NN expects the input samples to be independently distributed while consecutive states in RL and ADP are highly correlated. To address these challenges, we propose using off-policy updates. To do this, we save the current state and feasible action set for each vehicle $\forall_{i}\,(s_{t}^{i},{\cal F}_{t}^{i})$ during sample collection. Then, offline, we score the feasible actions using the value function and use the ILP create the best matching. Finally, we update the value function of the saved post-decision state with that of the generated next post-decision state. This is different from experience replay in standard Q-Learning because the state and transition functions are partly known to us and choosing the best action, in our case, involves solving an ILP. In addition to off-policy updates, we use standard approaches in DRL like using a target network and Double Q-Learning \cite{van2016deep}.

\subsubsection{Addressing the data scarcity:}
Neural Networks typically require millions of data points to be effective, even on simple arcade games~\cite{mnih2015human}. In our approach, we address this challenge in 3 ways:
\squishlist
\item Practically, we see that in the RMP, the biggest bottleneck in speed is in generating feasible actions. To address this, as noted above, we directly store the set of feasible actions instead of recomputing them for each update.
\item Secondly, we use the same Neural Network for the value function associated with each of the individual vehicles. This means that a single experience leads to multiple updates, one for each vehicle.
\item Finally, we use Prioritised Experience Replay \cite{schaul2015prioritized} to reuse existing experiences more effectively.
\squishend

\subsubsection{Practical simplifications:}
Finally, based on our domain knowledge, we introduce a set of practical simplifications that makes learning tractable: 
\squishlist
\item Instead of using one-hot representations for discrete locations, we create a low-dimensional embedding for each location by solving the proxy-problem of trying to estimate the travel times between these locations.
\item During training, we perform exploration by adding Gaussian noise to the predicted $V_i$ values \cite{plappert2017parameter}. This allows us to more delicately control the amount of randomness introduced into the training process than the standard $\epsilon$-greedy strategy.
\item We don't use the pre-decision state of all the other vehicles to calculate the value function for a given vehicle (as suggested in Section \ref{sect:approx}). Instead, we aggregate this information into the count of the number of nearby vehicles and provide this to the network, instead.
\squishend

\noindent The specifics of the neural network architecture and training can be found in the supplementary (Section \ref{sec:supp}).

\begin{figure*}
    \centering
    \includegraphics[width=0.99\textwidth,height=1.75in]{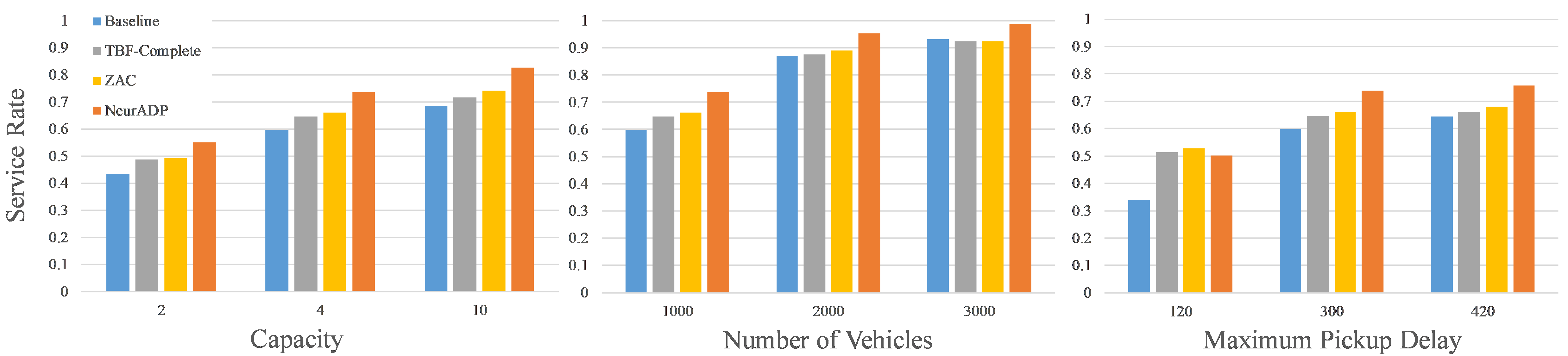}
    \caption{The three graphs benchmark our performance across 3 sets of parameter values - $c^{i}$, $\tau$ and $\left|{\cal R}\right|$ respectively (from left to right). In each case, we start with the prototypical configuration of $\tau$ = 300 seconds, $c^{i}$ = 4 and $\left|{\cal R}\right|$ = 1000 and vary the chosen parameter.}
	
    \label{fig:results}
\end{figure*}

\begin{figure}
    \centering
    \includegraphics[width=\hsize]{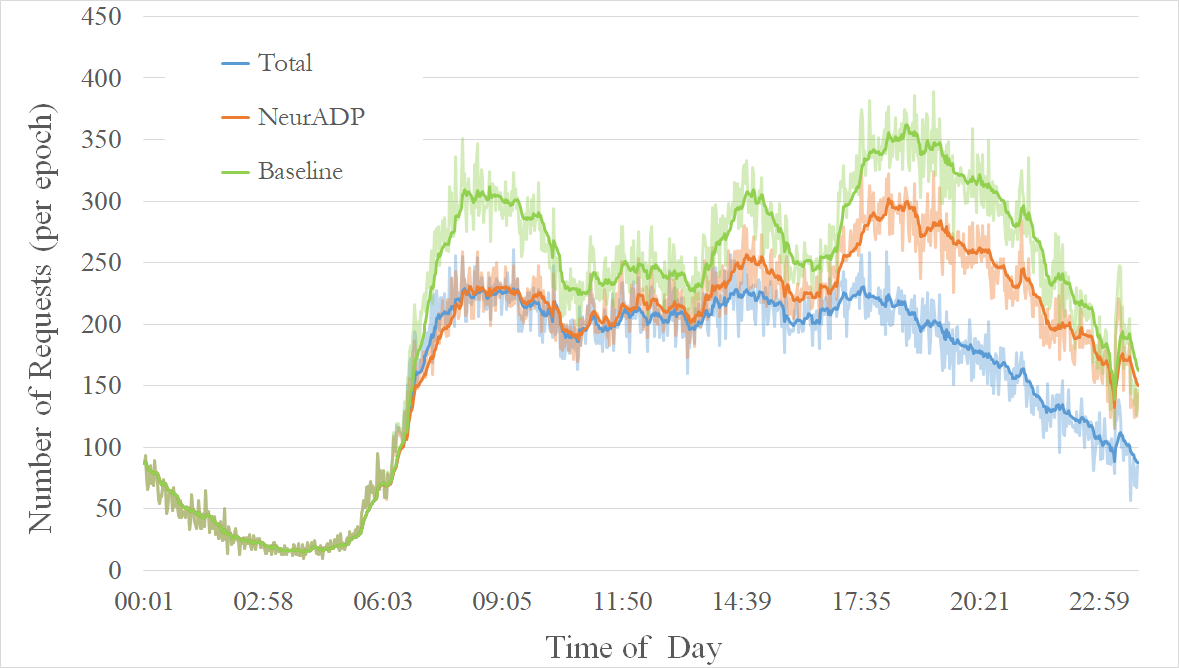}
	
    \caption{The graph compares the number of requests served as a function of time. The bold lines represent a moving average of the actual values (represented by the lighter lines). This graph corresponds to the configuration $\lambda$ = 300sec, $c^{i}$ = 10 and $\left|{\cal R}\right|$ = 1000 on 4 April 2016}

    \label{fig:reward_time}
\end{figure}

\section{Experiments}
The goal of the experiments is to compare the performance of our NeurADP approach to leading approaches for solving the RMP on a real-world dataset\cite{yellowtaxi} across different RMP parameter values. The metric we use to compare them is the service rate, i.e., the percentage of total requests served. Similar to \cite{alonso2017demand,lowalekarVJ19}, we vary the following parameters: the maximum allowed waiting time $\tau$ from 120 seconds to 420 seconds, the number of vehicles $\left|{\cal R}\right|$ from 1000 to 3000 and the capacity $c^{i}$ from 2 to 10. The value of maximum allowable detour delay $\lambda$ is taken as $2*\tau$. The decision epoch duration $\Delta$ is taken as 60 seconds. 

We compare NeurADP against the following algorithms: 
\squishlist
\item {\bf ZAC} -- ZAC algorithm by ~\cite{lowalekarVJ19}. 
\item {\bf TBF-Complete} -- Implementation of ~\cite{alonso2017demand} taken from ~\cite{lowalekarVJ19}. 
\item {\bf TBF-Heuristic (Baseline)} -- This is our implementation of Alonso et.al.'s~\cite{alonso2017demand} approach~\footnote{Please refer to the supplementary (Section \ref{sec:supp}) for a complete list of differences in the implementation.}.
\squishend

To disentangle the source of improvement in our approach, we introduce TBF-Heuristic which we refer to as the baseline. This uses a fast insertion method, that mirrors the implementation in NeurADP, to generate feasible actions. This is important because training requires a lot of samples and the key bottleneck in generating samples is the generation of feasible actions. While this process is completely parallelisable in theory, our limited academic computing resources do not allow us to leverage this. Therefore, comparing against this baseline allows us to measure the impact using future information has on solution quality. 

\noindent \textbf{Setup:} The experiments are conducted by taking the demand distribution from the publicly available New York Yellow Taxi Dataset \cite{yellowtaxi}. The experimental setup is similar to the setup used by \cite{alonso2017demand,lowalekarVJ19}. Street intersections are used as the set of locations ${\cal L}$. They are identified by taking the street network of the city from openstreetmap using osmnx with 'drive' network type \cite{boeing2017osmnx}. Nodes that do not have outgoing edges are removed, i.e., we take the largest strongly connected component of the network. The resulting network has 4373 locations (street intersections) and 9540 edges. Similar to earlier work \cite{alonso2017demand}, we only consider the street network of Manhattan as a majority ($\sim$75\%) of requests have both pickup and drop-off locations within it.

The real-world dataset contains data about past customer requests for taxis at different times of the day and different days of the week. From this dataset, we take the following fields: (1) Pickup and drop-off locations (latitude and longitude coordinates) - These locations are mapped to the nearest street intersection. (2) Pickup time - This time is converted to appropriate decision epoch based on the value of $\Delta$. The travel time on each road segment of the street network is taken as the daily mean travel time estimate computed using the method proposed in \cite{santi2014quantifying}. The dataset contains on an average 322714 requests in a day (on weekdays) and 19820 requests during peak hour.

We evaluate the approaches over 24 hours on different days starting at midnight and take the average value over 5 weekdays (4 - 8 April 2016) by running them with a single instance of initial random location of taxis~\footnote{All experiments are run on 24 core - 2.4GHz Intel Xeon E5-2650 processor and 256GB RAM. The algorithms are implemented in python and optimisation models are solved using CPLEX 12.8. The setup and code are available at https://github.com/sanketkshah/NeurADP-for-Ride-Pooling.}. NeurADP is trained using the data for 8 weekdays (23 March - 1 April 2016) and it is validated on 22 March 2016. For the experimental analysis, we consider that all vehicles have identical capacities. 

\noindent \textbf{Results:} We now compare the results of our approach, NeurADP, against past approaches. Figure \ref{fig:results} shows the comparison of service rate between NeurADP and existing approaches. As shown in the figure, NeurADP consistently beats all existing approaches across different parameters. Here are the key observations:
\squishlist
\item \textbf{Effect of changing the tolerance to delay, $\tau$:} NeurADP obtains a 16.07\% improvement over the baseline approach for $\tau=120$ seconds. The difference between the baseline and NeurADP decreases as $\tau$ increases. The lower value of $\tau$ makes it difficult for vehicles to accept new requests while satisfying the constraints for already accepted requests. Therefore, it is more important to consider future requests while making current assignments when $\tau$ is lower, leading to a larger improvement.
\item \textbf{Effect of changing the capacity, $c^{i}$:} NeurADP obtains a 14.03\% gain over baseline for capacity 10. The difference between the baseline and NeurADP increases as the capacity increases. This is because, for higher capacity vehicles, there is a larger scope for improvement if the future impact of making an assignment is taken into account.
\item \textbf{Effect of changing the number of vehicles, $|{\cal R}|$:} The difference between the baseline and NeurADP decreases as the number of vehicles increase. This is because, in the presence of a large number of vehicles, there will always be a vehicle that can serve the request. As a result, the quality of assignments plays a smaller role.
\squishend

\noindent For the specific case of $\tau=120$, NeurADP does not outperform ZAC and TBF-Complete because they use a more complex feasible action generation which allows them to leverage complex combinations of requests and their ordering. This becomes important as the delay constraints become stricter. If NeurADP is implemented with the complete search for feasible action generation, we expect it to outperform ZAC and TBF in this case as well.

We further analyse the improvements obtained by NeurADP over baseline by comparing the number of requests served by both approaches at each decision epoch. Figure \ref{fig:reward_time} shows the total number of requests available and the number of requests served by the baseline and our approach NeurADP at different decision epochs. As shown in the figure, initially at night time when the demand is low both approaches serve all available demand. During the transition period from low demand to high demand period, the baseline algorithm starts to greedily serve the available requests without considering future requests. On the other hand, NeurADP ignores some requests during this time to serve more requests in future. This allows NeurADP to serve more requests during peak time.

The approach can be executed in real-time settings. The average time taken to compute each batch assignment using NeurADP is less than 60 seconds (for all cases)~\footnote{60 seconds is the decision epoch duration considered in the experiments}.

These results indicate that using our approach can help ride-pooling platforms to better meet customer demand.

\section{Conclusion}
On-demand ride-pooling has become quite popular in transportation (through services like UberPool, LyftLine, etc.), food delivery (through services like FoodPanda, Deliveroo, etc.) and in logistics. This is a challenging problem as we have to assign each (empty or partially filled) vehicle to a group of requests. Due to the difficulty of making such assignments online, most existing work has focussed on myopic assignments that do not consider the future impact of assignments. Through a novel combination of approaches from ADP, ride-sharing and Deep Reinforcement Learning, we provide an offline-online approach that trains offline on past data and provides online assignments in real-time. Our approach, NeurADP improves the state of art by up to 16\% on a real dataset. To put this result in perspective, typically, an improvement of 1\% is considered a significant improvement on ToD for an entire city~\cite{xu2018large,lowalekarVJ19}. 
\section{Acknowledgements}
This research was supported by the Singapore Ministry of Education Academic Research Fund (AcRF) Tier 2 grant under research grant MOE2016-T2-1-174. This work was also partially supported by the Singapore National Research Foundation through the Singapore-MIT Alliance for Research and Technology (SMART) Centre for Future Urban Mobility (FM).

\section{Supplementary}
\label{sec:supp}
\begin{table*}
	\center
	\begin{tabular}{|l|l|}
		\hline
		\Tstrut\Bstrut
		\hfil\textbf{Alonso approach} & \hfil\textbf{Baseline}\\
		\hline
		\multicolumn{2}{|l|}{\Tstrut\Bstrut\hfil\textbf{Generation of Feasible Trips}} \\
		\hline
		\multicolumn{1}{|p{\columnwidth}|}{\Tstrut\Bstrut\raggedright 1. Generate RV graph by checking feasibility of each request with each vehicle. Keep only 30 closest vehicles for each request.} & \multicolumn{1}{p{\columnwidth}|}{1. Same}\\
		\multicolumn{1}{|p{\columnwidth}|}{\Tstrut\Bstrut\raggedright2. Perform exhaustive search for up to 4 requests (in the taxi currently and in the proposed trip). For more requests, check if the request can be inserted into the current vehicle path.} & \multicolumn{1}{p{\columnwidth}|}{2. Insert request into current vehicle path, irrespective of number of requests, for faster computation.}\\
		\multicolumn{1}{|p{\columnwidth}|}{\Tstrut\Bstrut\raggedright3. The exploration of feasible trips for a vehicle is stopped when a time limit of 0.2 seconds is reached.} & \multicolumn{1}{p{\columnwidth}|}{3. Exploration is stopped when feasibility constraints are evaluated 150 times. This is done to make the performance independent of the processing speed.}\\
		\hline
		\multicolumn{2}{|l|}{\Tstrut\Bstrut\hfil\textbf{Rebalancing Strategy}} \\
		\hline
		\multicolumn{1}{|p{\columnwidth}|}{\Tstrut\Bstrut\raggedright Number of vehicles rebalanced is min(unassigned vehicles, unassigned requests)} & \multicolumn{1}{p{\columnwidth}|}{All unassigned vehicles are rebalanced.}\\
		\hline
	\end{tabular}
	\caption{Differences between Baseline and Alonso Approach}
	\label{table:diff}
\end{table*}

\subsection{Neural Network and Training Specifics}
As inputs to this NN, we take the current location of the vehicle along with information about the remaining delay\footnote{The deadline to reach the location, according to quality constraints ${\cal C}$, minus the expected arrival time} and locations for the current requests that have been accepted. We order these according to the trajectory associated with them and feed them as inputs to an LSTM after an embedding layer. The embeddings for the locations are calculated separately and are the byproduct of a two-layer neural network that attempts to estimate the travel times between two locations.

Additionally, we add information about the current decision epoch, the number of vehicles in the vicinity of vehicle $i$ and the total number of requests that arrived in the epoch. Due to the constraint that a single request can only be assigned to a single vehicle, multiple agents compete for the same request. As a consequence, the value of being in a given state is dependent on the competition it faces from other agents when it is in that state. Adding the information about other taxis and the number of current requests stabilises learning significantly. These inputs are concatenated with the output of the LSTM from the previous paragraph and after 2 dense layers, used to predict the V-value. The loss considered is the mean squared error and it is minimised using the Adam optimiser using default initial parameters. We need to explore despite having determinstic transition and reward functions because the action space, in our model, is stochastic. 

This value function over individual vehicles is learned offline. When the approach is running online, we compute the assignment (of customer requests to vehicles) that maximizes the value function computed in the offline phase

\begin{table*}
	\center
	\scriptsize
	{
		\begin{tabular}{|l|l|l|l|l|l|l|l|}
			\hline
			~& \multicolumn{3}{|c|}{Parameters} &		\multicolumn{1}{|c|}{Baseline} &			\multicolumn{2}{|c|}{Our Approach}\\							
			\hline
			
			~& Number of & Pickup& Capacity & Requests &  Requests & Percentage \\
			~& Vehicle & Delay & ~& Served & Served & Improvement \\
			\hline
			Capacity & 1000 & 300 & 1 & $98581.4 \pm 1588.674$ & $121302.6 \pm  3008.226$ & $23.04816 \pm  2.719923$ \\
			\hline
			~& 1000 & 300 & 2 & $139497 \pm 3438.244$ & $177625.8 \pm 4791.103$ & $27.33306 \pm  2.312733$\\
			\hline
			~& 1000 & 300 & 4 & $192430 \pm 2687.394$ & $237534.2 \pm  7682.915$ & $23.43928 \pm  2.80972$\\
			\hline
			~& 1000 & 300 & 10 & $220737.4 \pm 5334.318$ & $266310 \pm 13789.16$ & $20.64562 \pm  3.910312$\\
			\hline
			
			Pickup Delay & 1000 & 120 & 4 & $109436 \pm 3309.862$ & $161257.8 \pm 2704.993$ & $47.35352 \pm 4.301165$\\      
			\hline
			
			~& 1000 & 300 & 4 & $192430 \pm 2687.394$ & $237534.2 \pm  7682.915$ & $23.43928 \pm 2.80972$\\
			\hline
			
			~& 1000 & 420 & 4 & $207396.6  \pm 5154.906$ & $243899.8 \pm 9279.68$ & $17.60067 \pm  1.821488$\\ 
			\hline
			
			Number of Vehicles & 1000 & 300 & 4 & $192430 \pm 2687.394$ & $237534.2 \pm 7682.915$ & $23.43928 \pm 2.80972$\\
			\hline
			
			~& 2000 & 300 & 4 & $280715.8 \pm 16320.16$ & $307306.6 \pm 15183.41$ & $9.472499 \pm 1.198605$\\
			\hline
			
			~& 3000 & 300 & 4 & $300122.6 \pm  16428.76$ & $318786.4 \pm 19384.14$ & $6.218725 \pm 1.300478$\\

			\hline
		\end{tabular}
	}
	\caption{Detailed Results}
	\label{table:results}
\end{table*}

\subsection{Details of Baseline algorithm}
There are some differences in our implementation of the approach by \cite{alonso2017demand}. We refer to our implementation as Baseline and the approach by Alonso et.al. as Alonso approach. The differences are highlighted in table \ref{table:diff}. The difference in generation of feasible trips is due to following practical considerations 
\begin{enumerate}
	\item Training time: To effectively train RL algorithms, we need a large number of experiences. To generate samples to train from, we must run our approach for some training days. Using  \cite{alonso2017demand}'s strategy for generating feasible trips takes significantly longer than the modification we propose. Given that our training time already takes multiple days, this is not viable. During test time, we maintain the same strategy to ensure coherence with what our value function is trained on.  
	\item Limitation due to academic computational resources: Our problem is completely parallelisable across different vehicles and so, in commercial set-ups, the consideration above would not stay relevant. In our case, however, we are bound by academic infrastructure.
\end{enumerate}

This is not a limitation for our approach. We expect the results of both the baseline and our approach improve proportionally if the feasible trips are generated as proposed in the paper by \cite{alonso2017demand}.

\subsubsection{Rebalancing}
Rebalancing empty vehicles has a significant impact on the number of requests served\cite{wallar2018vehicle}. Similar to \cite{alonso2017demand}, we perform a re-balancing of unassigned vehicles to high demand areas after each batch assignment. But unlike them we do not perform rebalancing by using only current unserved requests. This is because by using only current unserved requests for rebalancing, number of vehicles rebalanced will be minimum of unassigned vehicles and requests leaving majority of vehicles not being rebalanced in case of low demand scenarios. This means that vehicles that could be stuck in areas where requests are infrequent. Our approach differs from \cite{wallar2018vehicle} as we do not use the concept of 'regions' which are disjoint sets of locations. We work with individual locations, instead.

Therefore, we sample min(500,$|{\cal V}|$) requests from the number of requests seen so far and rebalance all vehicles to move to the areas of these sampled request by performing the optimization provided in table \ref{table:rebal_opt}. 

Let ${\cal V}^{t}_{u}$ denotes the set of unassigned vehicles at decision epoch $t$ and ${\cal D}^{t}$ denotes the set of sampled customer requests (as described above). $m_{ij}^{t}$ is a binary variable indicating that vehicle $i$ is moving towards customer request $j$. The objective of the linear optimization program is to minimize the sum of travel times. We use ${\cal T}(p_{i},o_{j})$ to denote the time taken to travel from initial location $p_{i}$ of vehicle $i$ to the origin $o_{j}$ of request $j$. 
Constraint \ref{cons:r1} ensures that each vehicle is assigned to exactly one request. Constraint \ref{cons:r2} ensures that each customer request is assigned to exactly $n_{j}^{t}$ vehicles where $n^{t}_{j} = \floor{\frac{|{\cal V}^{t}_{u}|}{500}}$ or $\ceil{\frac{|{\cal V}^{t}_{u}|}{500}}$ such that $\sum_{j \in {\cal D}^{t}} n^{t}_{j} =|{\cal V}^{t}_{u}|$.

\begin{table}
	\center
	\begin{tabular}{|r|}
		\hline
		\Tstrut
		\begin{minipage}{0.45\textwidth}
			\vspace{0.05in}
			\textbf{RebalanceVehicles(t):}
			{
				\small
				\begingroup
				\addtolength{\jot}{-2pt}
				\begin{align}
				\min \quad & \sum_{j \in {\cal D}^{t}} \sum_{i \in {\cal V}^{t}_{u}} {\cal T}(p_{i},o_{j})*m_{ij}^{t} \\
				subject ~ to \quad & \sum_{i \in {\cal V}^{t}_{u}} m_{ij}^{t} \leq n^{t}_{j} ::: \forall j \in {\cal D}^{t} \label{cons:r1}\\
				& \sum_{j \in {\cal D}^{t}} m_{ij}^{t} =  1 ::: \forall i \in {\cal V}^{t}_u \label{cons:r2}\\
				&0 \leq  m_{ij}^{t} \leq 1 ::: \forall i,j
				\end{align}
				\vspace{-8pt}
				\endgroup
			}
		\end{minipage} \\
		\hline
	\end{tabular}
	\caption{Optimization Formulation for Rebalancing unassigned vehicles}
	\label{table:rebal_opt}
\end{table}

\subsection{Detailed Results}
Table \ref{table:results} presents the average number of requests served by baseline and our algorithm over different days across different parameters.

\subsection{ADP for ToD Problems}
In order to provide a clear distinction between our contributions and the ADP methods employed in fleet optimization, we describe ADP for ToD problems~\cite{simao2009approximate}.  Each vehicle is represented using an attribute vector which captures the location and other information related to vehicle. The value function approximation used is linear 

\begin{align}
V(s_{t}^{a_t}) = \sum_{b \in {\cal B}} v_{t,b} \cdot R_{t,b}^{a_t} 
\end{align}

where ${\cal B}$ is the set of all possible attribute vectors and $R_{t,b}^{a}$ denotes the number of vehicles having attribute vector $b$ in post decision state. $v_{b}$ is the value of having an additional vehicle with attribute vector $b$ at time $t$. 

The value function approximation is improved by improving the estimate of $v_{b}$ values iteratively. Estimate of the parameter value at iteration $n$ is given by $v^n_{b}$:
\begin{align}
v_{b}^{n} = (1 - \alpha_{n-1})\cdot v_{b}^{n-1} + \alpha_{n-1} *\hat{v}_{b}
\end{align} 

where, $v_{b}^{n-1}$ represents the values at iteration $n-1$, $\hat{v}_{b}$ represents the dual values corresponding to the flow preservation constraint in the LP  and $\alpha_{n-1}$ is the step size. 

The pseudocode for ADP algorithm is shown in algorithm \ref{alg:basicadp}. 

%\noindent \textbf{\color{red} Sanket/Meghna: Here are the main comments in the section:\\
%1. Explain the LP dual values based update, as we are making a big deal about it.\\
%2. Algorithm updated according the notational changes above.\\
%3. List down the key drawbacks of ADP here.}

\begin{algorithm}
	\caption{ADP($N,T$)}
	{
		\small
		\begin{algorithmic}[1]
			\STATE{Initialization}
			\bindent
			\STATE{Initialize $V^{0}=0,$}
			\STATE{Set $n=1$}
			\eindent
			\FOR{$n < N$}
			\STATE{Initialize the state $s_{0}^{n}$ randomly.}
			\STATE{Choose a sample realization $\xi^{n}$ of exogenous information}
			\FOR{$0\leq  t \leq T$}
			\STATE{Solve the optimisation problem $v^{n} = \max_{a_{t} \in A_{t}} O(s_{t},a_{t})+ \gamma V^{n-1}(T^{a}(s_{t},a_{t}))$ to get actions $a_{t}^{n}$}
			\STATE{Update the value function $V^{n}$ using $v^{n}$} 
			\STATE{Update states $s_{t}^{a,n} = T^{a}(s_{t},a_{t}), s_{t+1}^{n} = T^{\xi}(s_{t}^{a},\xi_{t+1}^{n})$}
			\ENDFOR
			\ENDFOR
		\end{algorithmic}
	}
	\label{alg:basicadp}
\end{algorithm}
\bibliographystyle{aaai}
\bibliography{TaxiRef}

\end{document}